\documentstyle[12pt,a4wide,epsf]{article}

\begin{document}

\setcounter{footnote}{0}
\begin{titlepage}
\renewcommand{\thefootnote}{\fnsymbol{footnote}}
\makebox[2cm]{}\\[-1in]
\begin{flushright}
\begin{tabular}{l}
hep-ph/9506337\\
June 1995
\end{tabular}
\end{flushright}
\vskip0.4cm
\begin{center}
{\Large\bf
$|V_{cb}|$ from inclusive semileptonic B decays
}

\vspace{0.7cm}

Patricia Ball

\vspace{0.7cm}

{\em CERN, Theory Division, CH--1211 Gen\`{e}ve 23, Switzerland}

\vfill

{\bf Abstract\\[5pt]}
\parbox[t]{\textwidth}{
We determine $|V_{cb}|$ from inclusive semileptonic B
decays including resummation of supposedly large perturbative
corrections, which originate from the running of the strong coupling.
We argue that the low value of the BLM scale found previously for
inclusive decays is a manifestation of the renormalon divergence of
the perturbative series starting already in third order. A reliable
determination of $|V_{cb}|$ from inclusive decays is still
possible if one uses a short-distance b quark mass.
We find that using the $\overline{\rm MS}$ running mass significantly
reduces the perturbative coefficients already in low orders.
For a semileptonic branching ratio of $10.9\%$ we obtain
$|V_{cb}|(\tau_B/1.50\,{\rm ps})^{1/2}=0.041\pm 0.002\pm 0.002$. This work
was done in collaboration with M. Beneke and V.M.\ Braun.
}

\vspace*{2cm}

{\em
to appear in the Proceedings of the XXXth Rencontres de Moriond,\\
``QCD and High Energy Hadronic Interactions'',\\
Les Arcs, France, March 1995
}
\end{center}
\end{titlepage}
\renewcommand{\thefootnote}{\arabic{footnote}}
\setcounter{footnote}{0}
\newpage

Since it was shown that the decays of heavy hadrons can be described within
the framework of an operator product expansion (OPE) \cite{OPE}, the old
idea of extracting $|V_{cb}|$ from inclusive semileptonic B decays has
gained revived interest. A very welcome feature of the OPE is that it
expresses the hadronic decay rate as that of the parton model plus
non-perturbative
corrections, which are suppressed by two powers of the heavy quark mass:
\begin{equation}\label{eq:gamma}
\Gamma(B\to X_c e \nu) = \Gamma(b\to c e \nu)\,\left( 1 + \frac{\delta^{NP}}{
m_b^2}\right).
\end{equation}
The main efforts were first concentrated on determining
$\delta^{NP}/m_b^2$, which turned
out to be small and of the order of 5\%.
Subsequently, however, it was shown \cite{claims,pol} that the pole quark
mass, which
was habitually used in all analyses although it was known to be an ill-defined
quantity, suffers as manifestation of its ill-definedness from a renormalon
induced uncertainty of order $\Lambda_{\rm\scriptsize QCD}$, which would
generate
terms of order $1/m_b$ on the right-hand side of Eq.\ (\ref{eq:gamma}). As
a next step these uncertainties were shown to cancel against corresponding
ones in the perturbative corrections to $\Gamma(b\to c e \nu)$
\cite{claims,BBZ94}.
At the same time it turned out~\cite{BN94}, that the introduction of a
short-distance mass in (\ref{eq:gamma}), which {\em a priori} avoids all
renormalon uncertainties, shifts the value of $V_{cb}$ by $\sim 15\%$ at
one-loop level with respect to analyses using pole masses \cite{LS}.
Consequently, the question of higher order perturbative corrections to
the quark level decay moved in the center of interest and in Ref.\ \cite{LSW94}
the BLM prescription \cite{BRO83} was used to set the renormalization scale
in the lowest order radiative correction, which corresponds to taking into
account exactly the supposedly large correction in $\alpha_s^2\beta_0$,
$\beta_0$ being the lowest order coefficient of the QCD $\beta$-function.
The resulting ``optimum'' scale of order 500 MeV is rather small and
was interpreted as
an indication for the possible breakdown of perturbation theory in heavy quark
decays.

In Ref.\ \cite{BB94b}, cf.\ also \cite{Nnew}, we have generalized the
``standard'' BLM prescription to arbitrary order in $\alpha_s$ and applied
it to semileptonic b decays in Ref.\ \cite{Vcb}. The full series reads
\begin{equation}
\Gamma(b\to c e \nu)=\Gamma_0\Bigg\{1
 - C_F\frac{\alpha_s(m_b)}{\pi}g_0(a) \Big[1 +
\sum_{n=1}^\infty \tilde d_n(a)\alpha_s^n(m_b)\Big]\Bigg\}\,,
\end{equation}
where $a$ is the ratio of quark masses $m_c/m_b$ and $g_0(a)$ is the
one-loop correction first calculated in
\cite{Nir89}. Writing
\begin{equation}
       \tilde d_n(a) = \delta_n(a) + (-\beta_0)^n d_n(a)
\end{equation}
with $\beta_0 = -1/(4\pi)\{11-2/3\,N_f\}$, we can calculate with our method
both the $d_n(a)$ and the sum
\begin{eqnarray}
          && M_\infty^{b\to c}[a,-\beta_0\alpha_s(m_b)] \equiv
   1+\sum_{n=1}^\infty (-\beta_0)^n d_n(a)\alpha_s^n(m_b)\,,\\[-10pt]
&&\mbox{so that\ }\Gamma(b\to c e \nu)=\Gamma_0\Bigg\{1
 - C_F\frac{\alpha_s(m_b)}{\pi}g_0(a) M_\infty^{b\to c}
[a,-\beta_0\alpha_s(m_b)]\Bigg\}
\end{eqnarray}
in our approximation of neglecting the $\delta_n$.
In the BLM language the ``optimum'' scale is then $\mu_\infty^{b\to c}$,
defined as $\alpha_s(\mu_\infty^{b\to c}) \equiv \alpha_s(m_b)
M_\infty^{b\to c}[a,-\beta_0\alpha_s(m_b)]\,.$

Diagramatically, the $d_n$ can be obtained from
 the one-loop correction with the insertion of a chain of,
say, $i$ fermion loops into the gluon lines. These diagrams are
proportional to $N_f^i$, $N_f$ being the number of active quark flavours, and
thus proportional to $\beta_0^i$. Remarkably enough, these contributions can
be related to the one-loop correction $g_0(a,\lambda^2)$
 calculated with a {\em finite} gluon mass $\lambda$, such that \cite{BB94b}
($g_0(a)\equiv g_0(a,\lambda^2=0)$)
\begin{equation}\label{rBS}
g_0(a)(-\beta_0\alpha_s) M_\infty[a,-\beta_0\alpha_s] =
\int_0^\infty d\lambda^2\, \Phi(\lambda^2)\,g_0'(a,\lambda^2)
+[g_0(a,\lambda_L^2)-g_0(a)],
\end{equation}
where $\alpha_s =\alpha_s(\mu)$,
\begin{equation}\label{Phi}
\Phi(\lambda^2) = {}-\frac{1}{\pi}\arctan\left[\frac{-\beta_0\alpha_s\pi}
{1-\beta_0\alpha_s\ln(\lambda^2/\mu^2 e^C)}\right] -
\,\theta(-\lambda_L^2-\lambda^2)\,.
\end{equation}
Here $\lambda_L^2 =- \mu^2\exp[1/(\beta_0\alpha_s)-C]$
is the position of the Landau pole in the strong coupling and $C$ is a constant
characterizing the renormalization-scheme, $C=-5/3$ for the
$\overline{\rm MS}$-scheme and $C=0$ for the V-scheme.
In this talk I cannot give a detailed discussion of the asumptions underlying
Eq.\ (\ref{rBS}), but refer the  reader to the
corresponding sections in Ref.\ \cite{BB94b}. Still, two short comments are
appropriate.

First, note that the product
$\alpha_s(\mu) M_\infty[a,-\beta_0\alpha_s(\mu)]$ is explicitly
scale invariant, provided the coupling runs with leading-order accuracy.
The  result is also scheme-invariant, provided the couplings are
consistently related in the same BLM approximation,
that is by keeping only the terms with highest power in $N_f$.
Secondly, notice that the second term in (\ref{rBS}) involves the
radiative correction analytically continued to a negative squared gluon mass,
namely the position of the Landau pole, $\lambda^2_L<0$. The renormalon
divergence
of perturbation theory is reflected by non-analytic
terms in the expansion of $g_0(a,\lambda^2)$ at small $\lambda^2$ and
leads to an imaginary part in this continuation.
The size of the imaginary part (divided by $\pi$), $\delta M_\infty
\equiv 1/(\pi|\beta_0|\alpha_s)\,\mbox{Im}\,g_0(a,\lambda^2_L)$, yields
an estimate of the ultimate accuracy of perturbation theory, beyond
which it has to be complemented by non-perturbative corrections.
The real part of (\ref{rBS}) coincides with the
sum of the perturbative series defined by the principal value of the Borel
integral \cite{BB94b},
and the imaginary part of $g_0(a,\lambda^2_L)$ coincides with the
imaginary part of the Borel integral.
Numerically, we find
\begin{eqnarray}\label{result1}
    \Gamma(b\to u e \nu) &=& \Gamma_0\left\{
        1-2.41 \frac{\alpha_s(m_b)}{\pi}
\Big[1+0.75+0.67+0.70+0.87+1.27+\ldots\Big]\right\}
\nonumber\\
    &=& \Gamma_0\left\{
        1-2.41 \frac{\alpha_s(m_b)}{\pi}\Big[2.31\pm 0.62\Big]\right\}
\end{eqnarray}
using a b quark pole mass and
\begin{eqnarray}\label{result2}
    \Gamma(b\to u e \nu) &=& \overline{\Gamma_0}\left\{\!
        1+4.25 \frac{\alpha_s(\overline{m}_b)}{\pi}
\Big[1+0.604+0.159+0.073+0.032+\ldots\Big]\!\right\}
\nonumber\\
    &=& \overline{\Gamma_0}\left\{
        1+4.25 \frac{\alpha_s(\overline{m}_b)}{\pi}\Big[1.92\pm 0.01\Big]
\right\}
\end{eqnarray}
using the $\overline{\rm MS}$ mass $\overline{m}_b$. Obviously the
introduction of the
short-distance mass reduces the size of perturbative corrections considerably,
whereas the gross divergence of the series in (\ref{result1}) is caused by
the nearby $u=1/2$ renormalon that is bound to cancel the
corresponding one in the short-distance expansion of the pole mass.

\begin{figure}[t]
\epsfysize=2.4in
\centerline{\epsffile{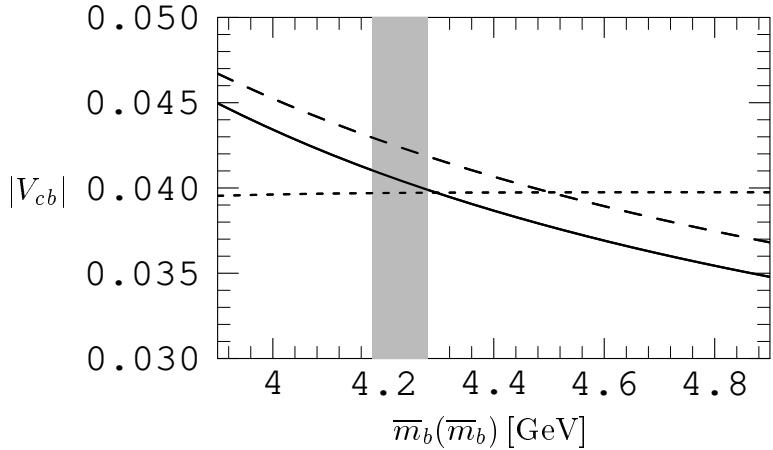}}
\caption[]{
The value of $|V_{cb}|$ extracted from the inclusive B meson
semileptonic decay rate after resumming
$\beta_0^n\alpha_s^{n+1}$ radiative corrections, shown as a function of the
$\overline{\rm MS}$ b quark mass for fixed $\lambda_1 = -0.5\,$GeV$^2$.
The solid and long-dashed curves show the
predictions obtained by using the $\overline{\rm MS}$ and OS scheme,
respectively. The central value coming from exclusive decays is shown by
short dashes and the shaded area gives the interval of b quark mass values
suggested by QCD sum rules. Experimental input:
$\tau_B = 1.5\,$ps, $B_{SL}=10.9\%$, $\alpha_s(m_Z)=0.117$.
}\label{fig:5}
\end{figure}
Let us now turn to the determination of $|V_{cb}|$. We still need to fix
$m_b$, $m_c$ and the non-perturbative correction $\delta^{NP}$ that enters
(\ref{eq:gamma}). We use the following value for the $\overline{\rm MS}$ b
quark mass suggested by QCD sum rules:\footnote{For
references and a critical discussion see \cite{Vcb}.}
$\overline{m}_b(\overline{m}_b) = (4.23 \pm 0.05)\,\mbox{\rm GeV.}$
In order to fix the c quark mass, we make use of the fact that the
{\em difference\/} between the pole masses of two heavy quarks is free
from many ambiguities intrinsic to the mass parameters themselves and
can be determined to a good accuracy from the expansion
\begin{equation}\label{relate-bc}
 m_b-m_c =m_B -m_D +\frac{1}{2}\left(\frac{1}{m_b}-\frac{1}{m_c}\right)
\Big[\lambda_1+3\lambda_2\Big]+O(\alpha_s/m,1/m^2),
\end{equation}
 where $m_B$ and $m_D$ are the B and D meson masses, respectively; $\lambda_2$
is given by $\lambda_2 \simeq 1/4\,(m^2_{B^*}-m^2_B) \simeq 0.12\,
\mbox{\rm GeV}^2,$
and $-\lambda_1/(2m_b)$ is the kinetic energy of a heavy quark inside a B
meson. For $\lambda_1$ an estimate is available from QCD sum rules
\cite{BABR94},
$\lambda_1=-(0.6\pm 0.1)\,$GeV$^2$.
As for the non-perturbative corrections, we just mention that they partly
depend on measurable quantities and partly on $\lambda_1$; in our analysis
we use the value $\delta^{NP} = -(1.05\pm 0.10)\,{\rm GeV}^2$.

In Fig.~1 we show $|V_{cb}|$ as function of the b quark mass. The solid line
shows the result obtained using $\overline{\rm MS}$ masses, the long-dashed
line is the result obtained for pole masses. The short dashes give $|V_{cb}|$
from exclusive decays. As compared to a corresponding analysis including only
$O(\alpha_s)$ terms, we find the dependence on the definition of the quark
masses to be considerably reduced.
The final value we thus extract is
\begin{equation}
 \left(\tau_B/{1.5\,{\rm ps}}\right)^{1/2}
|V_{cb}|_{incl} = 0.041\pm 0.002 \pm 0.002,
\end{equation}
where the first errors gives the theoretical uncertainty induced by the
errors in the values of $\overline{m}_b(\overline{m}_b)$ and $\lambda_1$
and the second
one comes from the uncertainty in the experimental branching ratio
\cite{CLEObr}. The full theoretical uncertainty inherent in our approach
is larger and in particular constituted by the uncalculated part of the
$\alpha_s^2$ corrections. It remains to be hoped that semileptonic heavy
quark decays do not behave different in that respect from other perturbative
expansions, where {\em a posteriori\/} the $\alpha_s^2\beta_0$ term indeed
turned out to be the dominant one, cf.\ the examples mentioned in Ref.\
\cite{BB94b}.

\end{document}